\documentstyle[aps,pre]{revtex}

\begin{document}

\draft

\title{Turbulent Photon Filamentation in Resonant Media}

\author{V.I. Yukalov}

\address{
Bogolubov Laboratory of Theoretical Physics \\
Joint Institute for Nuclear Research, Dubna 141980, Russia\\
and\\
Instituto de Fisica de S\~ao Carlos, Universidade de S\~ao Paulo\\
Caixa Postal 369, S\~ao Carlos, S\~ao Paulo 13560-970, Brazil}

\maketitle

\begin{abstract}

Optical turbulence is the term that is used being based on the analogy 
between hydrodynamic and optical equations.
The phenomenon of turbulent photon filamentation occurs in lasers and
other active optical media at high Fresnel numbers. A description of this
phenomenon is suggested. The solutions to evolution equations are
presented in the form of a bunch of filaments chaotically distributed in
space and having different radii. The probability distribution of patterns
is defined characterizing the probabilistic weight of different filaments.
The most probable filament radius and filament number are found, being in
good agreement with experiment.

\end{abstract}

\vskip 2cm


\vskip 1cm


\newpage

{\bf 1. Introduction}

\vskip 3mm

Different physical systems often display interesting similarity in their 
behaviour. This concerns, for example, hydrodynamic and optical systems.
There exists a well established analogy between the evolution equations
describing optical phenomena and hydrodynamic equations for compressible
viscous liquid [1--3]. The Fresnel number for optical systems plays the same
role as the Reynolds number for fluids. In optical systems, with increasing
Fresnel number, there appears chaotic behaviour similar to fluid turbulence
arising with increasing Reynolds number. Because of this analogy, one terms 
the {\it optical turbulence} [1--3] the chaotic phenomena occurring in
high-Fresnel-number optical systems.

For an optical cavity of the standard cylindrical shape of radius $R$ and
length $L$, the Fresnel number is $F\equiv\pi R^2/\lambda L$, with $\lambda$
being the optical wavelength. Physical processes accompanying the route
from the regular behaviour to the turbulent regime, arising with increasing
Fresnel numbers, are practically the same in different types of active optical
media, such as lasers and photorefractive crystals.

When Fresnel numbers are very small, $F\ll 1$, there exists the sole transverse
mode almost uniformly filling the optical cavity. At Fresnel numbers around
$F\sim 1$, several transverse modes emerge seen as bright spots in the
transverse cross-section. These spatial structures are regular both in space
and in time, since in space they form ordered geometric arrays, as polygons,
and because in time they are either stationary or oscillating periodically.
These transverse structures are imposed by the cavity geometry and correspond
to the empty-cavity Gauss-Laguerre modes. The appearance of such regular
geometric structures is well understood theoretically, with the theory being
in good agreement with experiments both for lasers [4--9] as well as for
photorefractive crystals [10--12]. For Fresnel numbers up to $F\approx 5$,
the number of regular modes is proportional to $F^2$.

The transition from the regular behaviour to a turbulent state occurs
around the Fresnel number $F\approx 10$. For Fresnel numbers $F>15$,
the arising spatial structures are principally different from those
defined by the empty-cavity Gauss-Laguerre modes. The related modal
description becomes no longer relevant and the boundary conditions do
not play the role any more. In the medium, there appears a large number
of filaments stretched along the cavity axis. In the transverse
cross-section, they are distributed chaotically. The correlation length is
much shorter than the mean distance between filaments, so that they radiate
independently of each other, aperiodically flashing in time. The number of
filaments, contrary to the regular regime, is proportional to $F$. This
chaotic filamentation has been observed in both photorefractive crystals
[10--12], e.g. in Bi$_{12}$SiO$_{20}$, as well as in many lasers, such as
Ne, Tl, Pb, N$_2$, and N$_2^+$ vapor lasers [13--17], CO$_2$ and dye lasers
[18--26]. The specific random properties of the observed filamentation, such
as chaotic distribution of filaments in space, the absence of correlations
between them, and their aperiodic flashing, are analogous to the properties
of turbulent fluids [1--3]. Because of this analogy of the chaotic photon
filamentation in high-Fresnel-number optical media with turbulent phenomena
in high-Reynolds-number viscous fluids, one commonly calls [18--26] the
former the optical turbulence or, because of the formation of bright
filaments with a high density of photons, it can be called the {\it
turbulent photon filamentation}.

Contrary to the low-Fresnel number case, with regular optical structures,
for which a good theoretical understanding has been achieved [4--12], for
the high-Fresnel-number regime of the turbulent photon filamentation, there
has been no persuasive theory suggested. This problem was addressed in Refs.
[27--30]. However, because of the complexity of the phenomenon, only some
oversimplified stationary models were considered. In the present paper, the
description of the phenomenon of turbulent photon filamentation is suggested,
based on realistic evolution equations. It has become possible to solve now
this problem owing to the recently developed method for treating nonlinear
differential equations, called the {\it scale separation approach} [31--33].

\vskip 5mm

{\bf 2. Turbulent photon filamentation}

\vskip 3mm

For the system of $N$ resonant atoms interacting with electromagnetic field,
we use the standard dipole approximation. The evolution equations to be
considered are those for the averages of atomic operators,
\begin{equation}
\label{1}
u(\vec r,t)\equiv \; < \sigma^-(\vec r,t)>\; , \qquad
s(\vec r,t)\equiv \; < \sigma^z(\vec r,t)> \; ,
\end{equation}
where $\sigma^-$ is a transition operator and $\sigma^z$ is the
population-difference operator. Equations for the quantities (1) can be
easily obtained following the usual way, by writing the Heisenberg equations
of motion, eliminating the field variables, and employing the standard
semiclassical and Born approximations. All this machinery is well known and
can be found, for instance, in the book [34]. In order to present resulting
equations in a compact form, it is convenient to introduce the notation
$f\equiv f_0+f_{rad}$ for an effective field acting on an atom. This field
consists of the term $f_0\equiv -i\vec d\cdot \vec E_0$, due to the cavity
seed field, and of the term
\begin{equation}
\label{2}
f_{rad}(\vec r,t) \equiv\;-\; \frac{3}{4}\; i\gamma\rho\; \int
\left [ \varphi(\vec r -\vec r\;')\; u(\vec r\;',t) -
\vec{e_d}^2 \;\varphi^*(\vec r -\vec r\;')\; u^*(\vec r\;',t)\right ]\;
d\vec r\; '
\end{equation}
describing the effective interaction of atoms through the common radiation
field. Here $\rho$ is the density of atoms, $\vec d\equiv d_0\vec e_d$ is
the transition dipole, and $\varphi(\vec r)\equiv\exp(ik_0|\vec r|)
/k_0|\vec r|$, with $k_0\equiv\omega_0/c$ and $\gamma\equiv 4k_0^3d_0^2/3$.
The cavity seed field selecting a longitudinal propagating mode, is taken, as
usual, in the form of plane waves, 
$$
\vec E_0 =\vec E_1 e^{i(kz-\omega t)} +
\vec E_1^* e^{-i(kz-\omega t)} \; .
$$ 
The resulting equations are
$$
\frac{du}{dt} = \; -\; (i\omega_0 +\gamma_2) u + sf \; , \qquad
\frac{ds}{dt} =\; -\; 2(u^* f + f^* u) -\gamma_1 (s-\zeta) \; ,
$$
\begin{equation}
\label{3}
\frac{d|u|^2}{dt} =\; - 2\gamma_2 |u|^2 + s (u^* f + f^* u) \; ,
\end{equation}
where $\omega_0$ is the transition frequency, $\gamma_1$ and $\gamma_2$ are
the longitudinal and transverse relaxation widths, respectively, and $\zeta$
is the pumping parameter.

We consider a cylindrical cavity, typical of lasers, with radius $R$ and
length $L$. For the case of high Fresnel numbers $F\gg 10$, the cavity can
house several transverse modes. Therefore, it is reasonable to look for
solutions of Eqs. (3) in the form of a bunch of $N_f$ filaments, so that
\begin{equation}
\label{4}
u(\vec r,t) =\; \sum_{n=1}^{N_f}\; u_n(\vec r,t)\Theta_n(x,y)\; e^{ikz} \; ,
\qquad
s(\vec r,t) =\; \sum_{n=1}^{N_f}\; s_n(\vec r,t)\Theta_n(x,y)\; ,
\end{equation}
where 
$$
\Theta_n(x,y)=\Theta(b_n^2-(x-x_n)^2+(y-y_n)^2)
$$ 
is a unit-step function; the pair $x_n$ and $y_n$ defines the axis of a 
filament; and $b_n$ is the radial distance from the filament axis, at 
which the solutions decrease by an order of magnitude. If the filament 
profile is approximately of normal law $\exp(-r^2/2r_n^2)$, with $r_n$ 
being the filament radius, then $b_n=\sqrt{2\ln 10}\; r_n$. The filaments 
do not interact with each other, because of which they cannot form a 
regular space structure, and the locations of their axes in the 
transverse cross-section are random. The absence of interactions between 
filaments is due to the fact that the kernel $\varphi(\vec r)$, playing 
the role of a Green function in the  effective interaction field (2), 
fastly oscillates and diminishes with increasing $r$. Let us stress that 
the presentation of solutions in the form of a bunch of filaments (4) does 
not exclude the possible case of a sole filament almost uniformly filling 
the whole sample. This is because the number of filaments and their radii 
will be defined in a self-consistent way. Substituting the form (4) into 
Eqs. (3), we obtain a system of equations for $u_n,\; s_n$, and $|u_n|^2$. 
To simplify these equations, we employ the mean-field approximation for 
the averages
\begin{equation}
\label{5}
u(t) \equiv\; \frac{1}{V_n}\; \int_{V_n}\; u_n(\vec r,t)\; d\vec r\; , \qquad
s(t) \equiv\; \frac{1}{V_n}\; \int_{V_n}\; s_n(\vec r,t)\; d\vec r\; ,
\end{equation}
and for the similarly defined $|u(t)|^2$, where the integration is over the
volume $V_n\equiv\pi b_n^2L$ of a cylinder enveloping a filament. For the
simplicity of notation, the index $n$ labelling filaments is dropped from the
left-hand sides of Eqs. (5). In this way, Eqs. (3) are transformed to the
system of equations
$$
\frac{du}{dt} =\; -(i\Omega +\Gamma) u - is\vec d\cdot\vec E_1\;
e^{-i\omega t} \; ,
$$
\begin{equation}
\label{6}
\frac{ds}{dt} =\; -4g\gamma_2|u|^2 -\gamma_1(s-\zeta) -4\;{\rm Im}\left (
u^*\vec d\cdot\vec E_1\; e^{-i\omega t}\right ) \; ,
\end{equation}
$$
\frac{d|u|^2}{dt} =\; -2\Gamma|u|^2 + 2s\;{\rm Im}\left ( u^*\vec d\cdot
\vec E_1\; e^{-i\omega t}\right )
$$
for the functions (5). Here the notations for the collective frequency
$\Omega\equiv\omega_0 +g'\gamma_2s$ and for the collective width
$\Gamma\equiv\gamma_2(1-gs)$ are used, with the effective coupling 
parameters
$$
g\equiv\;\frac{3\gamma\rho}{4\gamma_2V_n}\; \int_{V_n} \;
\frac{\sin[k_0|\vec r-\vec r\;'|-k(z-z')]}{k_0|\vec r-\vec r\;'|}\; d\vec r\;
d\vec r\;'\; ,
$$
\begin{equation}
\label{7}
g' \equiv\;\frac{3\gamma\rho}{4\gamma_2V_n}\; \int_{V_n} \;
\frac{\cos[k_0|\vec r-\vec r\;'|-k(z-z')]}{k_0|\vec r-\vec r\;'|}\; d\vec r\;
d\vec r\;'\; ,
\end{equation}
Equations (6) describe the evolution of a filament. These evolution equations
can be analysed by using the scale separation approach [31--33]. For applying
the latter, we, first, take into account the existence of usual small
parameters $\gamma_1/\omega_0\ll 1$ and $\gamma_2/\omega_0\ll 1$, and also
we consider the quasiresonance  case, when $|\Delta|/\omega_0\ll1$, with
the detuning $\Delta\equiv\omega-\omega_0$. Then the functional variable
$u$ in Eqs. (6) is to be classified as fast, as compared to the slow variables
$s$ and $|u|^2$. The latter are quasi-invariants for the fast variable.
Following the scale separation approach [31-33], we solve the equation for
the fast variable $u$, which is straightforward when $s$ is a quasi-invariant.
The found solution for $u$ has to be substituted into the right-hand sides of
the equations for the slow variables $s$ and $|u|^2$, with averaging these
right-hand sides over explicitly entering time of fast oscillations. This
procedure, complimented by the transformation
\begin{equation}
\label{8}
w\equiv |u|^2 -\alpha s^2 \; , \qquad
\alpha\equiv \; \frac{|\vec d\cdot\vec E_1|^2}{(\omega-\Omega)^2+\Gamma^2}\; ,
\end{equation}
where $\alpha$ is assumed to be small, $\alpha\ll 1$, leads to the equations
\begin{equation}
\label{9}
\frac{ds}{dt} =\; -4g\gamma_2w -\gamma_1(s-\zeta) \; , \qquad
\frac{dw}{dt} =\; -2\gamma_2(1-gs)w
\end{equation}
for the slow variables.

The properties of solutions to Eqs. (9) essentially depend on the value of
the effective coupling $g$ defined in Eq. (7). From this definition, it is
evident that the value of $g$ depends on the characteristics of the considered
filament, in particular, on the filament radius $r_n$ which enters Eq. (7)
through the radius $b_n$ of the enveloping cylinder. It is convenient to
introduce the dimensionless parameter $\beta\equiv\pi b_n^2/\lambda L$ and
to consider $g=g(\beta)$ as a function of this parameter in the domain
$0\leq\beta\leq F$. Thus, each filament can be characterized by its radius
or, equivalently, by the related parameter $\beta$. The bunch of filaments
parametrized by $\beta$ forms the solutions (4). The stability properties
of filaments with different values of $\beta$ are different, which can be
described by defining the probability distribution of patterns corresponding
to filaments with $\beta$ varying in the interval $[0,F]$. The latter can be
done as follows. Any system of evolution equations, like Eqs. (9), can always
be presented in the normal form $dy/dt=v$, in which $y=\{ y_i\}$, and
$v=\{ v_i\}$ is a velocity field. In the case of Eqs. (9), we have
$y_1\equiv s,\; y_2\equiv w$.

As is known from statistical mechanics, a probability $p$ is connected with
entropy $S$ by the relation $p\sim e^{-S}$. In the nonequilibrium case, it 
is more appropriate to count the entropy $S(t)$, which is a function of time,
from its initial value $S(0)$, thus, considering the entropy variation
$\Delta S(t)\equiv S(t)-S(0)$. The entropy is defined as the logarithm of an
elementary phase volume, $S(t)=\ln|\delta\Gamma(t)|$, where, in the 
nonequilibrium case, $\delta\Gamma(t)\equiv\prod_j\delta y_j(t)$. The 
elementary phase volume can be expressed as 
$\delta\Gamma(t)=\prod_i\sum_j M_{ij}(t)\delta y_j(0)$ through the 
elements $M_{ij}(t)\equiv\delta y_i(t)/\delta y_j(0)$ of the multiplier 
matrix $\hat M(t)= [M_{ij}(t)]$. Hence, the entropy variation is
$\Delta S(t)={\rm Tr}\hat L$, with the matrix $\hat L= [L_{ij}]$ being
composed of the elements $L_{ij}\equiv\ln|M_{ij}|$. By varying the 
evolution equations, one gets the equation $d\hat M/dt=\hat J\hat M$ for 
the multiplier matrix $\hat M(t)$, where $\hat J=[J_{ij}]$ is the 
Jacobian matrix with the elements $J_{ij}\equiv\delta v_i/\delta y_j$. In 
this way, the probability $p\sim e^{-\Delta S}$ acquires the form 
$p\sim\exp\{-{\rm Tr}\hat L\}$. From the equation for the multiplier 
matrix, one has the entropy variation $\Delta S(t)=\int_0^t\;K(t')dt'$, 
where $K\equiv{\rm Tr}\hat J$ called the contraction rate [35]. 
Therefore, for the probability distribution of patterns, labelled by a
parameter $\beta$, one obtains
\begin{equation}
\label{10}
p(\beta,t) =\; \frac{1}{Z(t)}\; \exp\left\{ -\; \int_0^t\;
K(\beta,t')\; dt'\right\} \; ,
\end{equation}
where 
\begin{equation}
\label{11}
Z(t) = \int \; \exp\left\{ -\int_0^t K(\beta,t')\; dt'\right\}\; d\beta
\end{equation}
is the normalization factor. As is clear, that pattern is preferred
over the other which possesses a higher probability weight. This implies,
because of the form (10), that the most preferable pattern is that for which
the {\it local contraction}
\begin{equation}
\label{12}
\Lambda(t) \equiv \frac{1}{t}\; \int_0^t \; K(t')\; dt'
\end{equation}
is minimal. At the initial stage of pattern formation, one may write 
$p\sim\exp\{-K(\beta,0)t\}$. Hence, the most probable pattern
to be formed is that corresponding to the minimal contraction rate
$K(\beta,0)$.

Considering the evolution equations (9), we keep in mind the standard way
of exciting laser media by means of a nonresonant pumping described by the
pumping parameter $\zeta>0$. No resonant fields are involved, so that
$s(0)<0$. It is easy to calculate the contraction rate $K(\beta,0)$
corresponding to Eqs. (9), which is 
\begin{equation}
\label{13}
K(\beta,0)=-\gamma_1- 2\gamma_2(1-gs_0)\; ,
\end{equation} 
where $s_0\equiv s(0)$. The minimum of this contraction
rate is provided by the maximum of $g=g(\beta)$. The maximum of $g(\beta)$,
with $g(\beta)$ given by Eq. (7), occurs at $\beta=0.96$. From here we find
the most probable filament radius
\begin{equation}
\label{14}
r_f=0.26\sqrt{\lambda L} \; .
\end{equation}
The most probable number of filaments can be evaluated from the normalization
integral 
$$
\frac{1}{V}\;\int s(\vec r,t)d\vec r=\zeta \; , 
$$
where the integration
runs over the whole volume of the sample, $V=\pi R^2L$. Considering this
normalization integral for the stage when the filaments have already been
formed and the population difference inside a filament of radius $r_f$ has
reached a value close to $\zeta$, we obtain
\begin{equation}
\label{15}
N_f =\left (\frac{R}{r_f}\right )^2 = 4.71 F\; .
\end{equation}
The number of filaments is proportional  to the Fresnel number $F$, which
is common for the turbulent photon filamentation.

Calculating the filament radius $r_f$ for the variety of experiments with 
different lasers, we find that the theoretical values of $r_f$ are in 
perfect agreement with all experimental data available. Thus, for Ne, Tl, 
Pb, N$_2$, and N$_2^+$ laser [13--17], we find $r_f\approx 0.01$ cm and 
for CO$_2$ and dye lasers [18--24], we have $r_f\approx 0.08$ cm and 
$r_f\approx 0.01$ cm, respectively. It is interesting that the 
coefficient $4.71$ in the dependence (15) for the number of filaments is 
also in good agreement with those experiments where it was measured [10--12].

\vskip 5mm

{\bf 3. Conclusion}

\vskip 3mm

The solution for the problem of turbulent photon
filamentation occurring in resonant media at high Fresnel numbers is
suggested. The consideration is based on realistic evolution equations
for resonant atoms under conditions typical of lasers. The solutions to
these equations have the form corresponding to a bunch of filaments with
different radii. The probability distribution of filament radii is found
self-consistently from the evolution equations. The results are in good
agreement with all experiments on the turbulent photon filamentation
in laser media [13--24]. The method presented in this paper can also be 
employed for considering other physical systems with arising 
spatio-temporal structures, when one confronts the so-called problem of 
pattern selection.

This work has been supported by the Bogolubov-Infeld Grant of the State
Agency for Atomic Energy, Poland and by a Grant of the S\~ao Paulo State
Research Foundation, Brazil.

\end{document}